# Identified particle correlation studies in central Au + Au collisions at $\sqrt{s_{NN}} = 200$ GeV

**Ying Guo** (*for STAR Collaboration*)

Wayne State University Department of Physics & Astronomy
666 West Hancock, Detroit MI 48201

We present correlations between strange baryons and mesons with charged hadrons in Au+Au and p+p collisions at $\sqrt{s_{NN}}$ =200 GeV for different $p_T$ ranges and centralities. We observe that for central Au+Au collisions, the large width of the back side correlations indicates a collapse of the jet structure. This is consistent with previous measurement of suppression of the back side correlation for unidentified charged hadrons. However, in the medium $p_T$ range the same side correlations show large AA/pp ratios and weak centrality dependencies which indicate different medium modification of the soft jet components. For different leading particle species we have found no significant difference for both same side and back side correlations. These results are comparable to PHENIX's measurements in a similar $p_T$ range [15].

## 1. Introduction

High $p_T$ particle azimuthal correlations have been used successfully in studying the jet suppression in central Au+Au collisions [1] [2] [3]. By using this method we can retrieve information about medium effects without the full reconstruction of the jet. It has been found that for unidentified charged hadrons in the $p_T$ ranges *4 GeV/c < $p_T$ trigger <6 GeV/c, 2 GeV/c< $p_T$ associated < 4 GeV/c* the back side jet correlations are suppressed in central Au+Au collisions [1]. These results show a strong medium modification effect: the jet suffers a significant energy loss inside the medium. Such an energy loss is in agreement with the partonic energy loss in a medium with high gluon density, therefore jet quenching is considered one of signatures of QGP. Many experimental results have shown the suppression of high $p_T$ particles in Au+Au collisions by measuring the nuclear modified factor $R_{AA}$ [4] [5]. Jet correlation studies can help us to understand the reasons for such suppressions; in order to gain more information of the medium and the interactions of the jets with the medium. Further more, the $R_{AA}$ and $R_{CP}$ measurements of different particle species have shown that the suppression has very different $p_T$ dependence for mesons and baryons [5] [6]. While the mesons are suppressed at relatively intermediate $p_T$ range (1.5~2.0 GeV/c), the bayrons exceed further into higher $p_T$ range ( 4~5 GeV/c) without experiencing strong suppression [9]. This led to the theory of possible different hadron production mechanisms such as parton recombination [7] [8] [9] instead of fragmentation of the jets in the intermediate $p_T$ range. Therefore particle identified jet correlations are very interesting since different particle production mechanisms are expected to contribute differently to the properties of the jet correlations. By tagging the jet with different particle species we can retrieve information about the flavor dependence and the particle dependence of the medium effects, energy loss and particle production mechanism etc. We can study the particle species dependence of the fragmentation function and its medium modification. This will eventually further help us to gain a more fundamental understanding of the strong interaction and QCD.

The particle reconstruction method used by STAR [10] allows us to measure strange bayrons ($\Lambda,\bar{\Lambda}$) and mesons ($K_0^s$) up to $p_T \sim 6\ GeV/c$ [11] [12]. The upper limit is determined by the statistics of the data and not the reconstruction method itself and this will be extended in the future RHIC runs. The reconstruction of the weakly decayed particle allows us to measure not only the bayron and meson differences in jet correlations but also differences between particle and anti particle, strange particle and non strange particle etc. The richness of the information can provide us a great handle on reasoning some of the effects such as suppression of the high $p_T$ strange particles, strangeness enhancement [13], and elliptic flow [14] etc, hence further complete our knowledge of the system created in heavy ion collisions at RHIC.

In this paper, we present two particle correlations with strange bayrons ($\Lambda,\bar{\Lambda}$) and mesons ($K_0^s$) as trigger particles and unidentified charge hadrons as associated particles in Au+Au collisions. The correlations are studied as a function of centrality and trigger particle species with two different $p_T$ cut sets. Two different models are applied to retrieve the parameters of the correlations for systematic studies. Results are compared to PHENIX's identified particle measurements.

## 2. Data Analysis and Results

The data presented in this paper are based on 2.6 million Au+Au events and ~10 million p+p events taken with the STAR detector. The central and the minbias trigger data are combined for the Au+Au analysis in order to improve the statistics. The leading particles such as $\Lambda,\bar{\Lambda}$ and $K_0^s$ are reconstructed from their decay daughters ($\Lambda \xrightarrow{B.R.\sim 63.9\%} p\pi^-, K_0^s \xrightarrow{B.R.\sim 68.6\%} \pi^+\pi^-$). This method allows us to identify strange bayrons and mesons to higher $p_T$ without explicit single track particle identification. Detailed information about tracking, efficiency correction, and V0 reconstruction is given in [11] [16]. The associated particles are unidentified charged hadrons. The correlation functions are calculated by simply plotting the difference between the azimuthal angle of leading particles and associated particles and then normalizing by the total number of the trigger particles. The two different $p_T$ cutsets presented in this paper are listed as the following. Only cut set B is applied to p+p data due to the limitation of the statistics.

Cut set A:    *1.7 GeV/c < $p_{T,associated}$ < 2.5 GeV/c < $p_{T,trigger}$ < 4.0 GeV/c*
Cut set B:    *1.5 GeV/c < $p_{T,associated}$ , $p_{T,trigger}$ < 3.0 GeV/c*

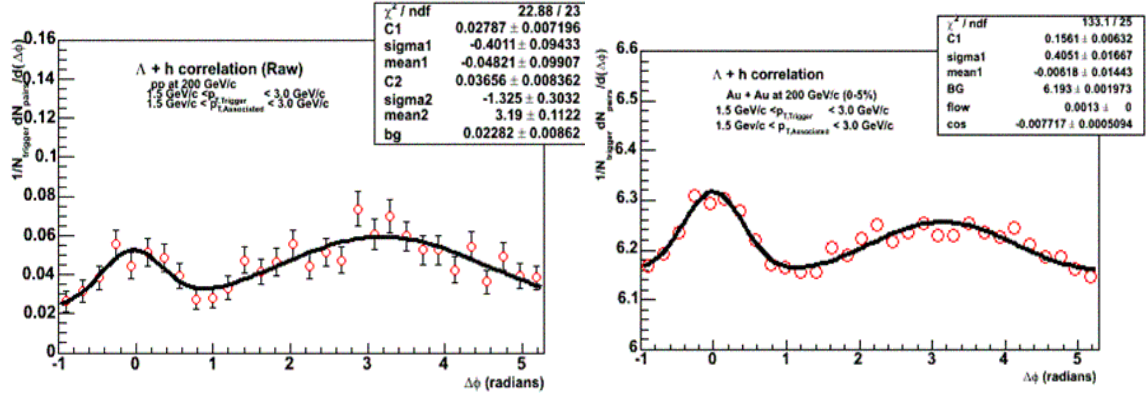

**Figure 1.** Sample correlation functions between Λ and charged hadrons for Au+Au and p+p

Figure 1 shows the correlations between Λ and charged hadrons for the central Au+Au and p+p collisions. Both correlations show same and back side jet like structures. We notice that the same side correlation is significantly enhanced for central Au+Au collisions compared to p+p collisions. We will discuss possible reasons in the later section. In Figure 1 the solid lines show the applied fit to the correlation functions. Two different functions are used for Au+Au and p+p due to their different compositions. Generally for Au+Au collisions the two particle correlation function in delta phi can have several components [1][17]. In the center of mass frame of the two scattering partons the final jet fragments will create a peak at zero (the same side correlation) and a peak at ***p*** (the back side correlation). These jet structures exceed the common flat statistics background reshaped by the elliptic flow. For p+p collisions there is no effect from elliptic flow. Generally both the same side correlation and the back side correlation can be fitted with a Gaussian. But for central Au+Au collisions because of the quenching effect, the back side can be treated as the momentum balance term due to the recoil of the quenched parton [18]. The formulas we have applied for the fits are:

for pp:
$$F(\Delta f) = a1 * \exp(-\frac{(\Delta f - \Delta f_1)^2}{2*s_1^2}) + a2 * \exp(-\frac{(\Delta f - \Delta f_2)^2}{2*s_2^2}) + bg \quad (1)$$

for Au+Au:

method A:  by assuming the back side is a Gaussian.
$$F(\Delta f) = a1 * \exp(-\frac{(\Delta f - \Delta f_1)^2}{2*s_1^2}) + a2 * \exp(-\frac{(\Delta f - \Delta f_2)^2}{2*s_2^2}) + bg(1 + 2.0 v_{2,trigger}^{flow} v_{2,asso}^{flow} \cos(2*\Delta f)) \quad (2)$$

method B: by assuming the back side is momentum balance cosine term

$$F(\Delta f) = a1 * \exp\left(-\frac{(\Delta f - \Delta f_1)^2}{2 * s_1^2}\right) + bg\left(C_{\cos}\cos(\Delta f) + 1 + 2.0 v_{2,trigger}^{flow} v_{2,asso}^{flow} \cos(2*\Delta f)\right) \quad (3)$$

Parameters are retrieved from different fits for comparison and calculations of the same side and back side yields. The benefit of using this method compared to the original $I_{AA}$ parameterization is that it avoids systematics due to the integration limits of $I_{AA}$. For Au+Au collisions five different centrality classes have been analyzed. ( *0-5%, 5-10%, 10-30%, 30-50% and 50-70%.* [1]) The elliptic flow parameter $v_2$ is extracted from the published data from previous measurement for charged hadrons and strange particles [16] [19]. The magnitude of the uncorrelated background can be calculated by using a convolution of the inclusive single particle spectra of the trigger and associated particles. [17] The centralities are scaled to the number of participants for nucleus-nucleus collisions based on a Glaubal model calculation [2]. Comparison of both the widths and the yields of the same side and back side jet structure are presented in figures 4-8. Figure 2 (left) and Figure 3 show the correlation of Lambdas with charged hadrons for cut set B for different centralities. The correlations are fitted with both method A (left) and B as shown in Figure 2 and 3, respectively. Apparently, both methods can describe the shapes of the correlation functions equally well.

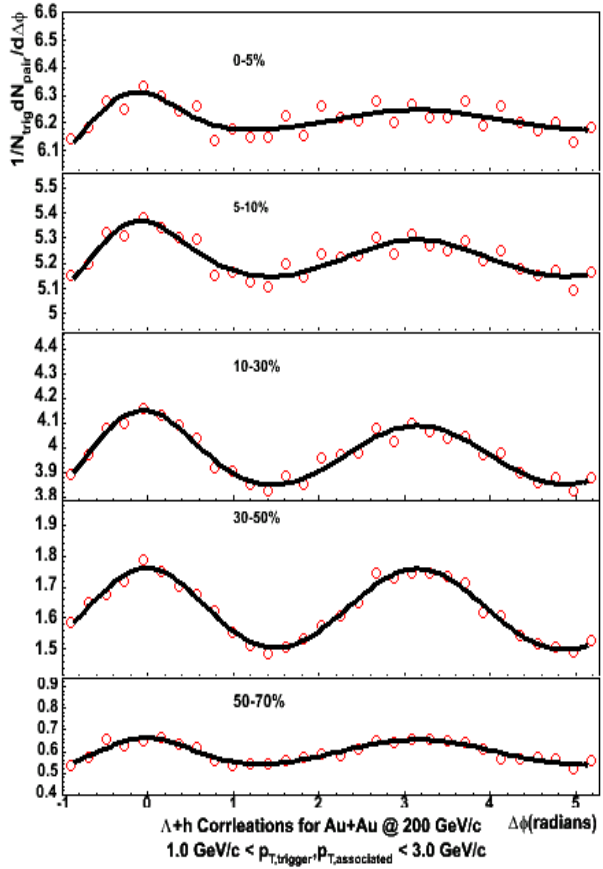 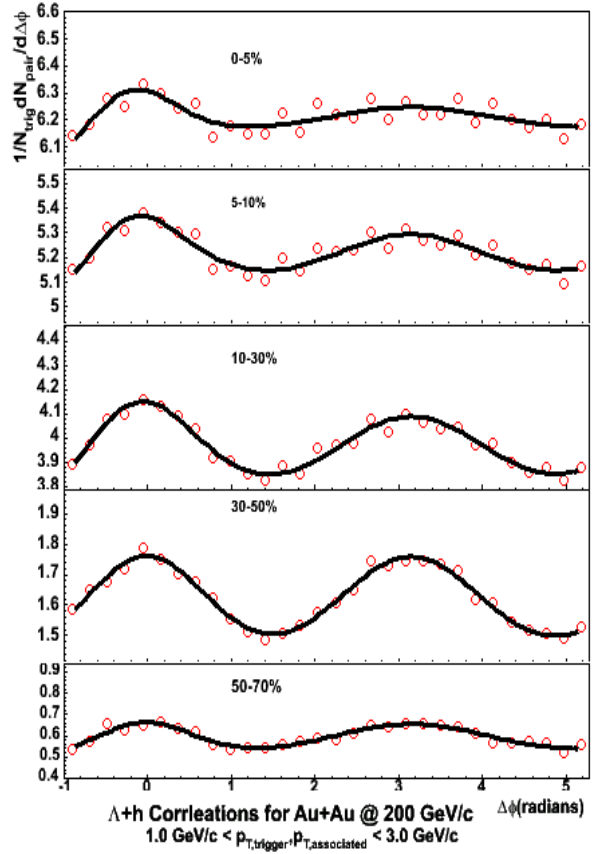

**Figure 2.** Correlations of $\Lambda$ with charged hadrons for cut set B for different centralities fitted with method A.

**Figure 3.** Correlations of $\Lambda$ with charged hadrons for cut set B for different centralities fitted with method B

Parameters are retrieved from both fits. Integration of the fit functions provides yields for same and back side jets. Figure 4 shows widths for the jet structure for both the same side correlation (open circle) and the back side correlation (open cross) as a function of the number of participants. Additionally, data from p+p is shown for comparison (open and solid triangle). For the same side correlations, the width shows a weak tendency from peripheral collisions to central collisions.

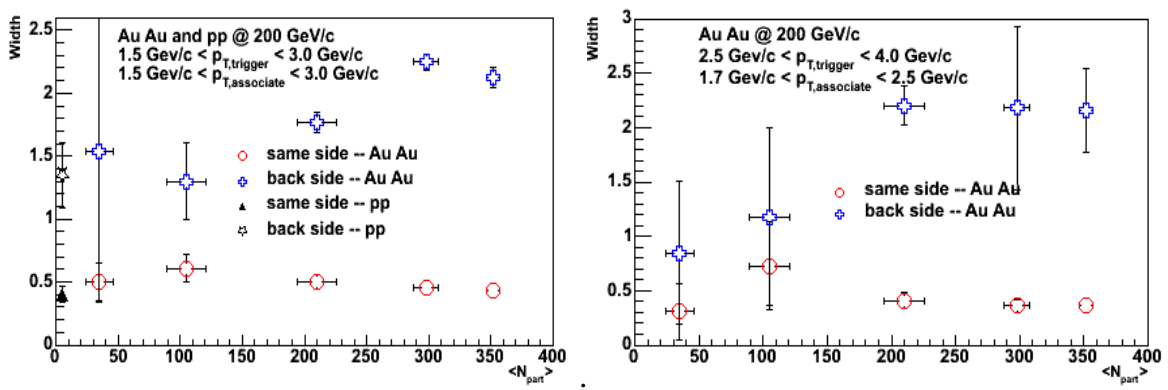

**Figure 4.** Width of the jet structure in $\Lambda$ - charged hadron correlations for Au+Au collisions at 200 GeV/c. (Errors are statistical only.)

Generally the magnitude is comparable to the p+p measurement. The systematic of the flow has to be further investigated to confirm this effect. For the back side correlation, the central Au+Au data set shows a very broad Gaussian width. In the most central collisions the back side width has exceeded $p/2$; this clearly shows the collapse of the jet structure. This is in agreement with previous unidentified charged hadron measurements which show the strong suppression of the back side jet structure in the most central collisions [1]. Due to the conservation of the transverse momentum of the quenched parton, a statistical cosine term can be applied to the back side correlations [18] in central collisions as an alternative to the simple Gaussian fit (Fitting method B). Such a term can be further analyzed to retrieve the total number of the particles that conserved the momentum [17]. Therefore we can gain additional information about the redistributions of the transverse energy of the quench parton inside the medium.

Figure 5 shows the same side and back side associated particle yields retrieved (from fitting method B) for different particle species as a function of the centrality of the Au+Au collisions. The different trigger particle species show similar trends, i.e. a weak centrality dependence with no significant difference except perhaps for the most central bin for the same side. We notice that if we form the Au+Au to

p+p ratio for the same side, the number becomes rather large ~*6.0* compared to *1.0* for the higher trigger particle $p_T$ range [1]. To understand this result we need to revisit our definition of the correlation function. The same side yield of the correlation should be proportional to the di-hadron production of a jet divided by the trigger-hadron production. The back side yield of the correlation should be proportional to single-hadron production of jet divided by the trigger-hadron production. In equations (4) and (5) we represent the same side and back side yields of the jet correlation with $N_{same}$ and $N_{back}$ respectively. We are using $P_1(p_T)$ to represent the single-hadron of the jet and $P_2(p_{T1}, p_{T2})$ to represent the di-hadron production of the jet. Generally because the probability of having a jet correlations when there is a trigger particle is less than 1, we are using $e_{same}$ to represent the probability of having a same side jet in the presence of the trigger particle and $e_{back}$ to represent the probability to have a back side jet. From (4) and (5) we can easily derive the back side to same side ratio as shown in equation (6).

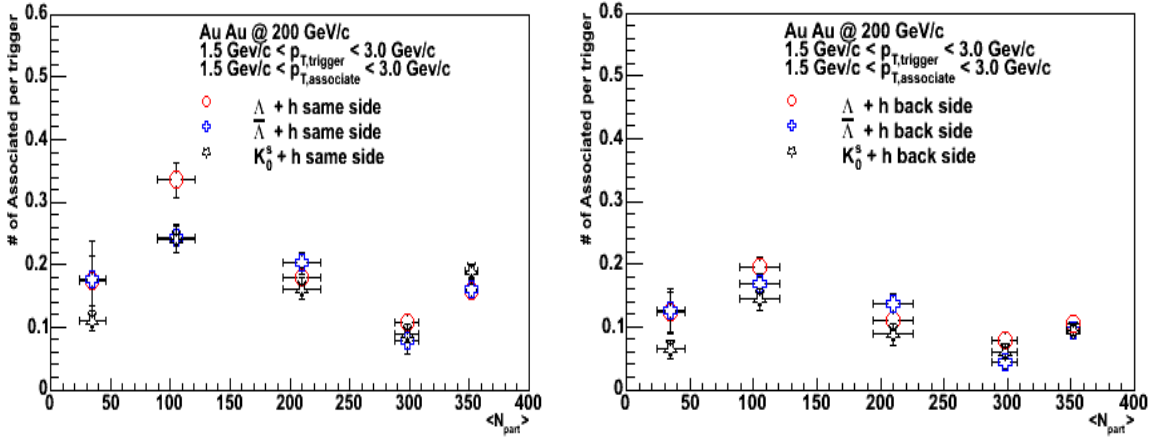

**Figure 5.** Yield of associated particles in same side and back side correlations for ($\Lambda, \bar{\Lambda}$) and ($K_0^s$) as a function of the number of participants

$$N_{same}(p_{T,trigger}) = e_{same} \frac{P_2(p_{T,associate}, p_{T,trigger})}{P_1(p_{T,trigger})} \qquad (4)$$

$$N_{back}(p_{T,trigger}) = e_{back} \frac{P_1(p_{T,associate})}{P_1(p_{T,trigger})} \qquad (5)$$

$$\frac{N_{back}}{N_{same}} = e \frac{P_1(p_{T,associate})}{P_2(p_{T,trigger}, p_{T,associate})} \quad \text{where} \quad e \equiv \frac{e_{back}}{e_{same}} \qquad (6)$$

For p+p collisions, when $e \sim 1.0$, this ratio is a function of trigger particle $p_T$ and associated particle $p_T$, and is generally greater than *1.0*. This is consistent with our observation for Lambda and charged hadron correlations for p+p collisions. For the central Au+Au collisions, the medium effect causes a decrease of $e$, therefore leads to the suppression of back side jet correlations. When we compare the same side correlation for Au+Au collisions to p+p, $e_{same}$ should generally not increase for central Au+Au because of the increasing thermal background. Therefore the larger Au+Au to p+p ratio should be due to the increase of the jet-like di-hadron production per jet, which could be the medium modification of the jet fragmentation or similar correlated particle production mechanisms, even on the same side. On the other hand the lack of strong differences in the correlations for different trigger particle species shows that in this $p_T$ range $e_{same}$ and $e_{back}$ are not sensitive to our measured particle species. In the framework of a simple thermalized parton recombination model one would expect a decrease of $e_{same}$ from peripheral collisions to central collisions since hadrons created by thermal partons should not be correlated. This should cause a weakening of the correlation function as a function of centrality in particular for bayrons which are enhanced in the intermediate $p_T$ range due to the parton coalescence [15]. We have not seen such effects in our correlation measurements. However, since the strength of the correlation is weak compared to the uncorrelated background, more systematic studies are required for $v2$ and background corrections to determine the centrality dependence.

We have done the same analysis for cutset B for strange bayrons $\Lambda$ and strange mesons $K_0^s$. As shown in Figure 6 the same side and back side yields are Comparable to similar measurements by PHENIX with bayrons and mesons as leading particles [15].

## 3. Summary

In this paper, we report high $p_T$ correlations with the identified strange bayron/meson as the trigger particles. The $p_T$ ranges for trigger hadrons and correlated hadrons are *2.5-4.0 GeV/c* and *1.7-2.5 GeV/c*, respectively. They all appear to have jet like structures. In these $p_T$ ranges, there is no significant difference between the correlations for different trigger particle species. For the central collisions in Au+Au, the large width of back side correlation indicates the collapse of the jet structure. This is consistent with the jet quenching picture and the previously measured suppression of the back side jets. We have also found the large Au+Au to p+p ratio for intermediate trigger $p_T$ range which indicates the medium modification of the fragmentation.

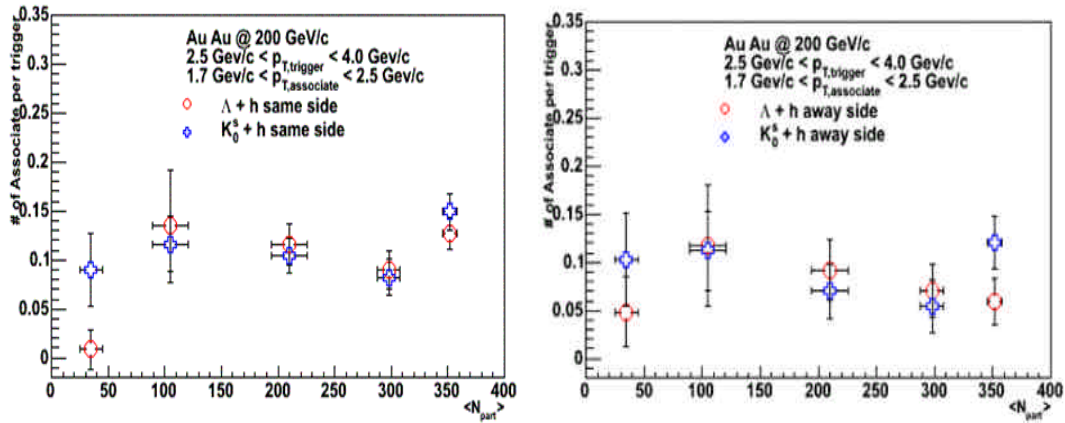

**Figure 6.** Same side yields for $\Lambda$ and $K_0^s$ as the function of the number of participants for different $p_T$ ranges

**Reference:**

[1]. C. Adler *et al.* Phys. Rev. Lett. 90, 082302 (2003)
[2]. J. Adams *et al.* Phys. Rev. Lett. 91, 072304 (2003)
[3]. F.Q. Wang, (STAR Collaberation), J. Phys. G30, S1299(2004)
[4]. C. Adler *et al.* Phys. Rev. Lett. 91, 172302 (2003)
[5]. C. Adler *et al.* Phys. Rev. Lett. 92, 052302 (2004)
[6]. S.S Adler *et al*. Phys. Rev. C 69, 034909(2004)
[7]. R.J. Fries et al: Phys. Rev. C 68 044902(2003)
[8]. D. Molnar , SA Voloshin, Phys. Rev. Lett. 91, 092301 (2003)
[9]. C. Adler *et al.* Phys. Rev. Lett. 92 052302(2004)
[10]. C Adler *et al* (STAR Collaboration) 2003 *Nucl. Instrum. Methods* A 499 624
[11]. C. Adler *et al* nucl-ex/0206008
[12]. C. Adler *et al.* Phys. Rev. Lett. 89, 092301(2002)
[13]. H. Caines, proceeding in this conference.
[14]. C. Adler *et al.* Phys. Rev. Lett. 87, 182301 (2001)
[15]. S.S Adler *et al*. nucl-ex/0408007
[16]. C. Adler *et al.* Phys. Rev. Lett. 89, 132301(2002)
[17]. Y. Guo (STAR Collaboration) hep-ex/0403018
[18]. N. Borghini *et al*. Phys. Rev. C 62, 034902(2000)
[19]. C. Adler *et al* nucl-ex/0407007